\documentclass[aps,pra,twocolumn,english,preprint,10pt,nofootinbib]{revtex4}
\usepackage[T1]{fontenc}
\usepackage[latin9]{inputenc}
\usepackage{babel}
\usepackage{nccmath}

\usepackage{graphicx,amsmath,amssymb,color}
\usepackage[normalem]{ulem}
\usepackage{amsfonts}
\usepackage[toc,page]{appendix}
\usepackage{hyperref}
\usepackage{latexsym}
\usepackage{amsfonts}
\usepackage{algpseudocode}
\usepackage{amsthm}
\usepackage{mathrsfs}
\usepackage{natbib}
\usepackage{color,verbatim}
\usepackage{psfrag}
\usepackage[svgnames]{xcolor}

\newcommand{\bra}[1]{\langle#1 |}
\newcommand{\ket}[1]{|#1 \rangle}

\newcommand{\be}{\begin{equation}}
\newcommand{\ee}{\end{equation}}
\newcommand{\bea}{\begin{eqnarray}}
\newcommand{\eea}{\end{eqnarray}}

\def\ket{\rangle}
\def\bra{\langle}
\def\l{\lambda}
\def\s{\sigma}
\def\g{\gamma}
\def\d{\delta}
\def\a{\alpha}
\def\b{\beta}
\def\m{\mu}
\def\n{\nu}

\def\conc{{\cal C}}

\preprint{IPPP/23/54}

\begin{document}

\title{Three-body Entanglement in Particle Decays}
\author{Kazuki Sakurai}
\email{kazuki.sakurai@fuw.edu.pl}
\address{Institute of Theoretical Physics, Faculty of Physics,
University of Warsaw, ul. Pasteura 5, PL-02-093 Warsaw, Poland}

\author{Michael Spannowsky}
\email{michael.spannowsky@durham.ac.uk}
\address{Institute for Particle Physics Phenomenology, Department of Physics, Durham University, Durham DH1 3LE, U.K.}

\begin{abstract}
Quantum entanglement has long served as a foundational pillar in understanding quantum mechanics, with a predominant focus on two-particle systems. We extend the study of entanglement into the realm of three-body decays, offering a more intricate understanding of quantum correlations. We introduce a novel approach for three-particle systems by utilising the principles of entanglement monotone concurrence and the monogamy property. Our findings highlight the potential of studying deviations from the Standard Model and emphasise its significance in particle phenomenology. This work paves the way for new insights into particle physics through multi-particle quantum entanglement, particularly in decays of heavy fermions and hadrons.
\end{abstract}

\maketitle

%\today % added for drafting, remove in final version

\section{Introduction}
\label{sec:intro}

The study of quantum entanglement has been a cornerstone of quantum mechanics, providing profound insights into the non-local correlations between quantum systems \cite{aspelmeyer, nonloc}. Historically, much of the focus has been on two-particle entanglement, particularly in the context of the Bell inequalities \cite{PhysicsPhysiqueFizika,PhysRevLett.28.938}. However, as the field of quantum mechanics evolves, it becomes imperative to explore more complex systems, specifically, the realm of multi-particle entanglement.

Recent research has delved into the intricacies of two-particle entanglement in the context of particle physics. Bi-partite systems have been explored for top-anti-top quarks \cite{Afik:2020onf,Fabbrichesi:2021npl,Severi:2021cnj,Aguilar-Saavedra:2022uye,Dong:2023xiw,Aoude:2022imd}, the Higgs boson \cite{Barr:2021zcp, Altakach:2022ywa,Aguilar-Saavedra:2022mpg}, gauge bosons \cite{Fabbrichesi:2022ovb,Barr:2022wyq,Ashby-Pickering:2022umy,Fabbrichesi:2023cev} and leptons \cite{Cervera-Lierta:2017tdt}, revealing that the quantum information properties of their spin states at proton colliders are accessible in current data. Furthermore, some studies emphasised the importance of quantum observables in probing the underlying dynamics of quantum systems \cite{Beane:2018oxh,Low:2021ufv,Carena:2023vjc}.

Given the advancements in studying two-particle entanglement, reflected by the large number of entanglement measures for bipartite systems \cite{Bennett:1995tk,Bennett:1996gf,Vedral:1997qn,Hill:1997pfa,Horodecki:2009zz}, extending this research to three-particle systems is a natural yet surprisingly underexplored advancement. The study of three-particle entanglement presents a richer tapestry of quantum correlations and offers the potential to uncover new insights into the fundamental nature of quantum mechanics. Moreover, extending the Bell inequality tests to three particles can provide a more robust framework for testing the foundational principles of quantum mechanics and exploring potential deviations from the predictions of the Standard Model.

By building on the entanglement monotone {\it concurrence}
 and the {\it monogamy} property, we propose an approach to extend entanglement to three particles, charting a course for future explorations in the realm of multi-particle quantum entanglement in particle phenomenology. 
Extending the concept of entanglement to three particles will bolster its applicability to uncharted territories in particle phenomenology, e.g. the decay of heavy fermions and hadrons\footnote{In many high-energy processes, three-body decays are favoured phenomenologically over two-body decays, either because of kinematic reasons or improved observability over backgrounds. 
Amongst others, those include top decays, $t \to b f \bar f'$, and the Higgs boson decay, $H \to Z \mu^+ \mu^-$.} This exploration into hadron decays offers a fresh perspective and a deeper understanding of the intricate quantum interactions within these systems. 

Additionally, the introduction of the three-particle entanglement measure presents a new observable. This novel measure holds significant promise in aiding the search for unknown heavy resonances and potentially discovering new physics. 
The entanglement property is based on the fundamental interaction of the process.
Thus, for a comprehensive assessment of the expected three-particle entanglement in three-body decays, we calculate its value for the effective Lorentz structures generated by (pseudo)scalars, (pseudo)vectors and (pseudo)tensors exchanges, respectively.

In the following discussion, we assume these spin states do not decohere by hadronisation or interactions with the environment before the measurement. For this assumption to be warranted, the spin-decoherence time scale needs to be significantly longer than the lifetime of the decaying resonance. This is usually the case for electroweak-scale and many hadronic resonances. For example, the spin-decoherence time scale in top quark decays is  ${\cal O}(m_t/\Lambda^2_{\rm QCD})$, while its lifetime is many orders of magnitude shorter, i.e. $1/\Gamma_t$ with $\Gamma_t \simeq 1.4$ GeV.

\section{Definition of Entanglement}
\label{sec:entangle}

Entanglement can be quantified 
by a class of non-negative functions called 
{\it entanglement monotones} \cite{Horodecki:2009zz, Chitambar_2019},
whose values do not increase 
%can only decrease 
under local operations and classical communication (LOCC).  
A particularly convenient entanglement monotone is  {\it concurrence} \cite{Hill:1997pfa,Wootters1998}.
For a mixed state $\rho$ of two qubits the concurrence is defined as
\be
{\cal C}[\rho] = {\rm max}(0, \eta_1 - \eta_2 - \eta_3 - \eta_4)\,\in\, [0, 1],
\label{defC}
\ee 
where $\eta_i$ ($\eta_i > \eta_j$ for $i < j$) are the eigenvalues of
the matrix $R \equiv \sqrt{ \sqrt{\rho} \tilde \rho \sqrt{\rho} }$ with 
$\tilde \rho \equiv (\s_y \otimes \s_y) \rho^* (\s_y \otimes \s_y)$.
%${\cal C}[\rho] \in [0, 1]$. 
For separable states ${\cal C} = 0$, while
${\cal C} = 1$ for maximally entangled states. 
For a pure state of two qubits, $| \psi \ket \in {\cal H}_A \otimes {\cal H}_B$, the concurrence can be computed more straightforwardly as 
\be
{\cal C}[| \psi \ket] = 
\sqrt{ 2 (1 - {\rm Tr} \rho^2_B) } \,,
\label{defCpure}
\ee
where $\rho_B$ is the reduced density operator of subsystem $B$ obtained by tracing over subsystem $A$:
$\rho_B \equiv {\rm Tr}_A( | \psi \ket \bra \psi | )$.

For a three-qubit state, $| \Psi \ket \in {\cal H}_i \otimes {\cal H}_j \otimes {\cal H}_k$,
one can consider two types of entanglement. 
One is an entanglement between two individual particles, say between $i$ and $j$.
This entanglement can be computed by first tracing out subsystem $k$ and use formula \eqref{defC}:
\be
{\cal C}_{ij} = {\cal C}[\rho_{ij}],~~~~
\rho_{ij} = {\rm Tr}_k ( | \Psi \ket \bra \Psi | )
\label{Cij}~.
\ee
Another type is an entanglement between one particle and the rest of the system,
known as {\it one-to-other} bipartite entanglement. 
The concurrence between $i$ and the composite subsystem ($kj$) 
can be computed using Eq.\ \eqref{defCpure}:
\be
\conc_{i(kj)} = \sqrt{2 (1 - {\rm Tr} \rho^2_{kj})},
~~~~
\rho_{kj} = {\rm Tr}_i ( | \Psi \ket \bra \Psi | )~.
\label{Ci(jk)}
\ee
Here we used ``Qubit Power'' of
the Schmidt theorem \cite{Schmidt1907} (see also e.g.\ \cite{Xie2021}) and applied the two-qubit formula Eq.\ \eqref{defCpure} to a three-qubit state $| \Psi \ket$.

The entanglement between $i$ and subsystem $(kj)$
cannot be freely shared between $i$-$j$ and $i$-$k$.
Namely, there is a trade-off between $i$'s entanglements with $j$ and $k$.
This property, called {\it monogamy}, is one of the most fundamental traits of entanglement
and formulated by the Coffman-Kundu-Wootters (CKW) monogamy inequality \cite{Coffman:1999jd,Osborne2006}:
\be
{\cal C}^2_{i(kj)} \ge {\cal C}^2_{ij} + {\cal C}^2_{ik} \,.
\label{monogamy}
\ee

For multipartite systems, one can define a so-called
{\it genuine multipartite entanglement} (GME)
\cite{Dur:2000zz,Ma_2011,Xie2021}.
A good GME measure should (1) vanish for all product and biseparable states,
(2) be positive for all non-biseparable states,
and (3) not increase under LOCC.
Recently, a GME measure satisfying all these criteria has been found for three-qubit states  
\cite{Jin2023}.
It corresponds to the area of the {\it concurrence triangle}, whose three sides are given by 
the three one-to-other bipartite entanglements:
\be
F_3 \,=\, 
\left[ \tfrac{16}{3} Q ( Q - {\cal C}_{1(23)}) ( Q - {\cal C}_{2(13)}) ( Q - {\cal C}_{3(12)}) \right]^{\frac{1}{2}}\,,
\ee
with $Q = \frac{1}{2}[{\cal C}_{1(23)} + {\cal C}_{2(13)} + {\cal C}_{3(12)}]$.
With this definition, $F_3$ takes values between 0 and 1.

\section{Entanglement in 3-body decays}
\label{sec:entangle_3B}

\begin{figure}[!t]
\centering
\includegraphics[scale=.28]{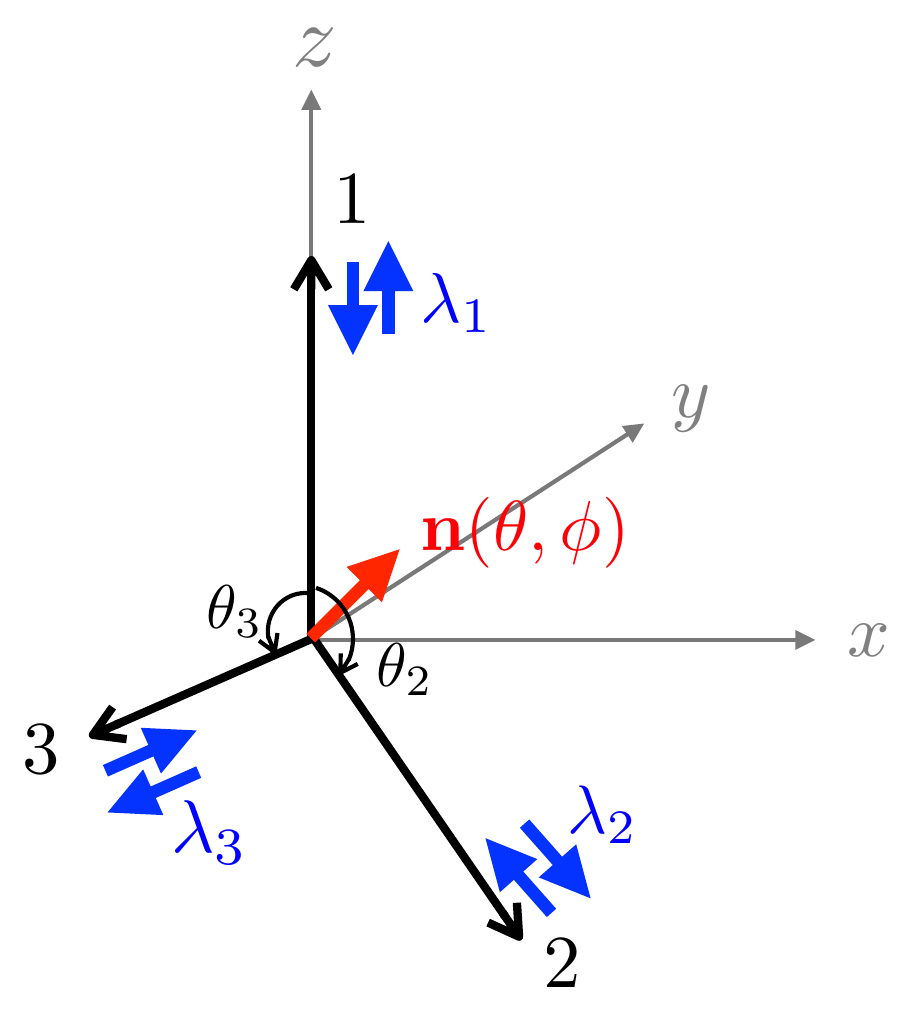}
\caption{The momentum and spin configuration in the coordinate system.
% The momentum of 1 is fixed to the $z$-direction.
% At the rest frame of 0,
% the decay plane is aligned with the $x$-$z$ plane 
% and the two opening angles, 1-2 and 1-3, are given by
% $\theta_2$ and $\theta_3$, respectively,
% with $0 \le \theta_2, \theta_3 \le \pi$ and
% $\pi \le \theta_2 + \theta_3 \le 2 \pi$.
% The spin direction of 0 is given by
% ${\bf n}=(s_\theta c_\phi, s_\theta s_\phi, c_\theta)$.
\label{fig:angle}
}
\end{figure}

We consider a 3-body decay $0 \to 123$
and assume all particles 
are distinguishable and
have spin-1/2.
We analyse the entanglement 
of the spin degrees of freedom (d.o.f.) of the final state particles 1, 2 and 3
at a given phase-space point (${\bf p}_1$, ${\bf p}_2$, ${\bf p}_3$).\footnote{As a related topic, the entanglement in a
orthopositronium decay into three photons 
%(qutrit $\to$ three identical bosonic qubits)
has been studied in Ref.\ \cite{Hiesmayr:2017xgx}.
}
To parametrise the phase space of the final state we boost into the rest frame of 
the initial particle 0 and take the $z$ axis in the direction of ${\bf p}_1$.
The $x$ and $y$ axes are chosen such that 
the $y$ axis is perpendicular to the decay plane and the ${\bf p}_2$ has a positive $x$-component. 
The opening angles $1 \to 2$ and $1 \to 3$ are denoted by $\theta_2$ and $\theta_3$ ($0 \le \theta_2, \theta_3 \le \pi$), respectively.
We represent the spin polarisation ${\bf n}$ of the initial particle 0 by the polar and azimuthal angles, $\theta$ and $\phi$, respectively (see Fig.\ \ref{fig:angle}).

We choose the spin quantisation axis of each final state particle in the momentum direction of that particle.  In this case, the eigenvalues of the spin (multiplied by 2) are called {\it helicity} and denoted by $\l_i = \pm 1$ ($i = 1,2,3$). 

For a given set of interactions, 
the quantum field theory framework 
lets us calculate the transition matrix element
(helicity amplitude)
\be
{\cal M}^{\bf n}_{\l_1, \l_2, \l_3}
= \bra \l_1, \l_2, \l_3 | {\bf n} \ket\,,
\ee
where the momentum labels are suppressed.
The initial state $| {\bf n} \ket$
is expanded by the final states as
\be
| {\bf n} \ket = \sum_{\l_1, \l_2, \l_3}
{\cal M}^{\bf n}_{\l_1, \l_2, \l_3}
| \l_1, \l_2, \l_3 \ket
\,+\, \cdots 
\,.
\label{expand}
\ee 
The ellipsis represents 
final states of other phase-space points 
and other decay modes.
Focusing on the spin d.o.f.\ 
one can describe the final spin state as
\be
| \Psi \ket = \frac{1}{\cal N} 
\sum_{\l_1, \l_2, \l_3}
{\cal M}^{\bf n}_{\l_1, \l_2, \l_3}
| \l_1, \l_2, \l_3 \ket \,,
\ee
where ${\cal N} = ( \sum_{\l_1, \l_2, \l_3}
|{\cal M}^{\bf n}_{\l_1, \l_2, \l_3}|^2 )^{1/2}$
is the normalisation constant. 
In general, this is an entangled pure state of three qubits.

The 2-particle entanglement 
and one-to-other entanglement 
defined in Eqs.~\eqref{Cij} and \eqref{Ci(jk)}
can be readily calculated, respectively.

In the following, we assume, for simplicity, that the final state particles are massless while 
the extension to the massive case is straightforward. In general, there are $16$ non-redundant Lorentz structures formed from from bilinear combinations of Dirac spinors $\bar{\psi} \Gamma \psi$  with
\begin{equation}
\Gamma = \left \{ \mathbb{I},\gamma^5, \gamma^\mu, \gamma^\mu \gamma^5, \sigma^{\mu \nu} \right \}~,
\end{equation}
where $\g^\m$ is the Dirac $\gamma$ matrices, $\g^5 \equiv i \g^0 \g^1 \g^2 \g^3$
and $\sigma^{\m \n} \equiv \frac{i}{2} [\g^\m, \g^\n]$.

As a 3-body decay $0\to 123$ of one fermion into three fermions requires two bilinears, 256 Lorentz structures form a complete basis. Instead, we will focus on the matrix elements and Lorentz structures induced by the exchange of (pseudo)scalars, (pseudo)vectors and (pseudo)tensors.

\subsection{Scalar and Pseudoscalar Interaction}

We consider the effective interaction operator
\be
[ \bar \psi_1 (c_S + i c_A \g_5) \psi_0 ]
[ \bar \psi_3 (d_S + i d_A \g_5) \psi_2 ]\,,
\ee
where $c_S, c_A, d_S, d_A \in {\mathbb R}$ are coupling constants.
We also define $c = c_S + i c_A$ and $d = d_S + i d_A$
and take $|c| = |d| = 1$ as we are not interested in the overall scale of the amplitude. 
For given phase-space point $(\theta_2, \theta_3)$ and the initial spin $(\theta, \phi)$,
the matrix element of $0 \to 123$ can be calculated as
\bea
&& {\cal M}^{\bf n}_{\l_1,\l_2,\l_3}
\propto
%& \propto & 
%[ \bar u_1 (c_S + i c_A \g_5) u_0 ]
%[ \bar u_3 (d_S + i d_A \g_5) v_2 ]
%\nonumber \\
%&=&
2 \sqrt{2 m p_1 p_2 p_3} \cdot 
s \tfrac{\theta_2 + \theta_3}{2} 
\nonumber \\
&& \Big[\,
-\, c d \cdot \d^-_{\l_1} \d^-_{\l_2} \d^-_{\l_3}   
\cdot e^{i \phi}  s \tfrac{\theta}{2} 
\,+\,
 c d^* \cdot \d^-_{\l_1} \d^+_{\l_2} \d^+_{\l_3}   
\cdot e^{i \phi}  s \tfrac{\theta}{2} 
\nonumber \\
&&-\,
 c^* d \cdot \d^+_{\l_1} \d^-_{\l_2} \d^-_{\l_3}   
\cdot c \tfrac{\theta}{2} 
\,+\,
 c^* d^* \cdot \d^+_{\l_1} \d^+_{\l_2} \d^+_{\l_3}   
\cdot c \tfrac{\theta}{2} 
\Big] \,,
%\nonumber \\
%
\eea
where shorthand notations 
$c \a = \cos \a$ and $s \a = \sin \a$ are used.
This corresponds to the spin state
\bea
| \Psi \ket &\,=\,& M_{LL} |--- \ket + M_{LR} |-++ \ket \nonumber \\
&+& M_{RL} |+-- \ket + M_{RR} |+++ \ket,
\label{psi_state}
\eea
with
$M_{LL} = - \frac{c d}{\sqrt{2}} \cdot e^{i \phi}  s \tfrac{\theta}{2}$,
$M_{LR} = \frac{c d^*}{\sqrt{2}} \cdot e^{i \phi}  s \tfrac{\theta}{2}$,
$M_{RL} = - \frac{c^* d}{\sqrt{2}} \cdot c \tfrac{\theta}{2}$
and
$M_{RR} = \frac{c^* d^*
}{\sqrt{2}} \cdot c \tfrac{\theta}{2}$.
We see that this is a biseparable state
\bea
| \Psi \ket &=& 
\big[  c e^{i \phi} s \tfrac{\theta}{2}
 | - \ket_1 +
 c^* c \tfrac{\theta}{2}
| + \ket_1 
\big] \otimes
 \nonumber \\
 &&
\tfrac{1}{\sqrt{2}}
 \big[ d^* | ++ \ket_{23} - d | -- \ket_{23} 
\big]\,.
\eea
Therefore, 1 is entangled neither with 2, 3 nor (23):
\be
{\cal C}_{12} = {\cal C}_{13} = {\cal C}_{1(23)} = 0,
\ee
while 2 and 3 are maximally entangled:
\be
{\cal C}_{23} = 1 \,.
\ee
The monogamy inequality \eqref{monogamy} implies 
2 and 3 must also be maximally entangled with the rest of the system:
\be
{\cal C}_{2(13)} = {\cal C}_{3(12)} = 1,
\ee
which can be explicitly checked from the formula \eqref{Ci(jk)}.
Because the state is biseparable, 
the GME measure vanishes
\be
F_3 = 0\,.
\ee

\subsection{Vector and Axialvector Interaction}

We next consider the vector interaction
\be
[ \bar \psi_1 \gamma_\m (c_L P_L + c_R P_R) \psi_0 ]
[ \bar \psi_3 \gamma^\m (d_L P_L + d_R P_R) \psi_2 ]\,,
\ee
with $P_{R/L} \equiv (1 \pm \g_5)/2$ and
$c_L, c_R, d_L, d_R \in {\mathbb R}$.
The matrix element is found as
\bea
&& {\cal M}^{\bf n}_{\l_1,\l_2,\l_3} 
%\nonumber \\
%&& = [ \bar u_1 \g^\m (c_L P_L + c_R P_R) u_0 ]
%[ \bar u_3 \g_\m (d_L P_L + d_R P_R) v_2 ]
%\nonumber \\
%&& 
\propto
4 \sqrt{2 m p_1 p_2 p_3} 
\nonumber \\ 
&&~~~~\,
\Big[\, \d_{\l_1}^- \d_{\l_2}^+ \d_{\l_3}^- \cdot 
c_L d_L  
s \tfrac{\theta_3}{2} \left[
c \tfrac{\theta}{2} c \tfrac{\theta_2}{2} 
+ e^{i \phi} s \tfrac{\theta}{2} s \tfrac{\theta_2}{2}
\right] 
\nonumber \\
&&~~-
\d_{\l_1}^- \d_{\l_2}^- \d_{\l_3}^+
\cdot c_L d_R
s \tfrac{\theta_2}{2} \left[
 c \tfrac{\theta}{2} c \tfrac{\theta_3}{2} 
- e^{i \phi} s \tfrac{\theta}{2} s \tfrac{\theta_3}{2}
\right]
\nonumber \\
&&~~+
\d_{\l_1}^+ \d_{\l_2}^+ \d_{\l_3}^-
\cdot c_R d_L
s \tfrac{\theta_2}{2} \left[
 c \tfrac{\theta}{2} s \tfrac{\theta_3}{2} 
+ e^{i \phi} s \tfrac{\theta}{2} c \tfrac{\theta_3}{2}
\right]
\nonumber \\
&&~~+
\d_{\l_1}^+ \d_{\l_2}^- \d_{\l_3}^+
\cdot c_R d_R
s \tfrac{\theta_3}{2} \left[ 
 c \tfrac{\theta}{2} s \tfrac{\theta_2}{2} 
- e^{i \phi}  s \tfrac{\theta}{2} c \tfrac{\theta_2}{2}
\right] \Big] \,,
\eea
corresponding to the state
\bea
| \Psi \ket &=&
M_{LL} |-+- \ket + M_{LR} |--+ \ket
\nonumber \\
&+& M_{RL} |++- \ket + M_{RR} |+-+ \ket\,,
\label{psi_state}
\eea
with
$M_{LL} = {\cal M}^{\bf n}_{-+-}/{\cal N}$,
$M_{LR} = {\cal M}^{\bf n}_{--+}/{\cal N}$,
$M_{RL} = {\cal M}^{\bf n}_{++-}/{\cal N}$
and
$M_{RR} = {\cal M}^{\bf n}_{+-+}/{\cal N}$.
From explicit calculations, we find 
\bea
{\cal C}_{12} = {\cal C}_{13} = 0\,,~~
{\cal C}_{23} =
2 | M_{LL} M_{LR}^* + M_{RL} M_{RR}^* |\,.
\label{Cij_vec}
\eea
For one-to-other entanglements, we obtain
\bea
{\cal C}_{2(13)} &=& {\cal C}_{3(12)} 
\nonumber \\
&=&
2 \sqrt{\left( | M_{LL} |^2 + | M_{RL} |^2 \right)
\left( | M_{LR} |^2 + | M_{RR} |^2 \right)},
\nonumber \\
{\cal C}_{1(23)} &=&
2 \big| M_{RR} M_{LL} - M_{LR} M_{RL} \big|
\,.
\label{Cijk_vec}
\eea
Since all three $C_{i(jk)}$ are non-vanishing in general,  
the GME measure $F_3$ is also non-vanishing in that case.
$M_{XY} \propto c_X d_Y$ ($X,Y = L,R$)
and we see that both ${\cal C}_{23}$, ${\cal C}_{1(23)}$ and $F_3$
are proportional to $|c_L c_R d_L d_R|$.
One the other hand, ${\cal C}_{2(13)}$
and ${\cal C}_{3(12)}$ vanish 
only if $|c_L c_R| = |d_L d_R| = 0$. 

To discuss the momogamy relation,
we define the monogamy measure as
\be
M_i = {\cal C}^2_{i(jk)} - [ {\cal C}^2_{ij
} + {\cal C}^2_{ik} ],
\ee
for $i \neq j \neq k \neq i$.
The CKW monogamy inequalities are expressed by
$M_i \ge 0$ for $i=1,2,3$.
From Eqs.\ \eqref{Cij_vec} and \eqref{Cijk_vec},
one can show 
${\cal C}^2_{23} = {\cal C}^2_{2(13)} - {\cal C}^2_{1(23)}$.
In the vector interaction, we therefore have
\be
M_1 \,=\, M_2 \,=\, M_3 \,=\,
{\cal C}^2_{1(23)} \, \ge \, 0\,.
\ee

\subsection{Tensor and Pseudotensor Interaction}

We consider the tensor interaction
\be
[ \bar \psi_1 (c_M + i c_E \g_5) \s^{\m \n} \psi_0 ]
[ \bar \psi_3 (d_M + i d_E \g_5) \s_{\m \n} \psi_2 ]\,,
\ee
with $c_M, c_E, d_M, d_E \in {\mathbb R}$.
As in the scalar case, we define $c = c_M + i c_E$ and $d = d_M + i d_E$ and take $|c| = |d| = 1$.
This operator is equivalent to 
$ \a [\bar \psi_1 \s^{\m \n} \psi_0 ]
[ \bar \psi_3 \s_{\m \n} \psi_2 ]
- \frac{\b}{2} 
\epsilon^{\m \n \rho \s}
[ \bar \psi_1 \s_{\m \n} \psi_0 ]
[ \bar \psi_3 \s_{\rho \s} \psi_2 ]
$
with $\a = c_M d_M - c_E d_E$ and $\b = c_M d_E + c_E d_M$. 
The $0 \to 123$ matrix element is given by
\bea
&& {\cal M}^{\bf n}_{\l_1,\l_2,\l_3} %&=& [ \bar u_1 (c_M + i c_E \g_5) \s^{\m \n} u_0 ]
%[ \bar u_3 (d_M + i d_E \g_5) \s_{\m \n} v_2 ]
%\nonumber \\
\propto
- 8 \sqrt{2 m p_1 p_2 p_3} 
%\cdot \big\{
\nonumber \\
&& \Big[ \, c^* d^* \cdot \d^+_{\l_1} \d^+_{\l_2} \d^+_{\l_3}   
\cdot 
[ 2 e^{i \phi} s \tfrac{\theta}{2} s \tfrac{\theta_2}{2} s \tfrac{\theta_3}{2}
-
c \tfrac{\theta}{2} s \tfrac{\theta_2 - \theta_3}{2} ] 
\nonumber \\
&&+\,
 c d \cdot \d^-_{\l_1} \d^-_{\l_2} \d^-_{\l_3}   
\cdot 
[e^{i \phi} s \tfrac{\theta}{2} s \tfrac{\theta_2 - \theta_3}{2} 
+
2 c \tfrac{\theta}{2} 
s \tfrac{\theta_2}{2} s \tfrac{\theta_3}{2}
]
\Big] \,,
%\nonumber \\
%
\label{M_tens}
\eea
implying the spin quantum state
\be
| \Psi \ket \,=\,
M_{R} |+++ \ket + M_{L} |--- \ket
\,,
\label{psi_state3}
\ee
with
$M_{R} = {\cal M}^{\bf n}_{+++}/{\cal N}$
and
$M_{L} = {\cal M}^{\bf n}_{---}/{\cal N}$.
This state interpolates between the separable states,
$|+++\ket$ and $|---\ket$,
and the maximally entangled 
Greenberger Horne Zeilinger state,
$|{\rm GHZ} \ket = (|+++\ket + |---\ket)/\sqrt{2}$
\cite{GHZ}.
For the tensor interaction, there are no entanglements between 
two individual particles
\be
{\cal C}_{12} =
{\cal C}_{13} =
{\cal C}_{23} = 0 \,,
\label{Cij_tens}
\ee
while one-to-other entanglements are universal
\be
{\cal C}_{1(23)} = {\cal C}_{2(13)} = {\cal C}_{3(12)} =
2 |M_R M_L|\,.
\ee
The GME measure in this case is
\be
F_3 = 4 |M_R M_L|^2\,.
\ee
The monogamy inequalities are trivially satisfied 
because no entanglement of one-to-other is shared by the individual pairs (c.f.\ Eq.\ \eqref{Cij_tens}).  

As in the scalar case all entanglement measures are independent of the phases of $c$ and $d$.

\subsection{Numerical Results}

\begin{figure}[!t]
\centering
\includegraphics[scale=.32]{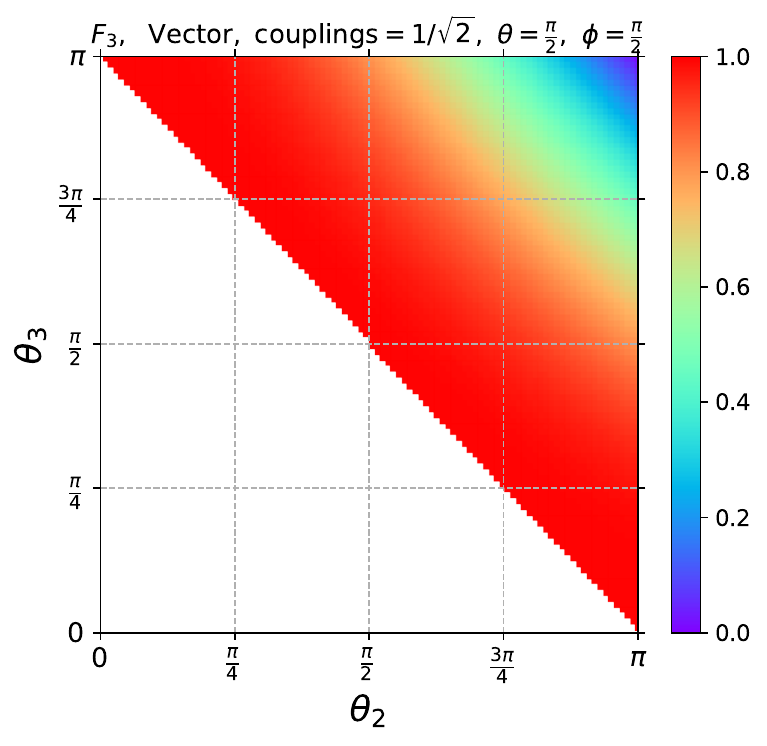}
\includegraphics[scale=.32]
{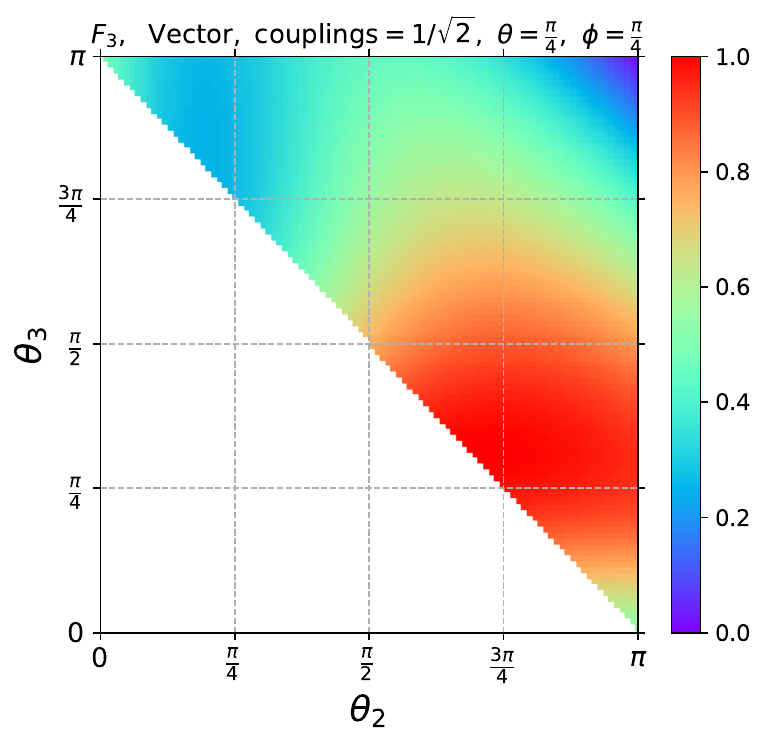}
\includegraphics[scale=.32]{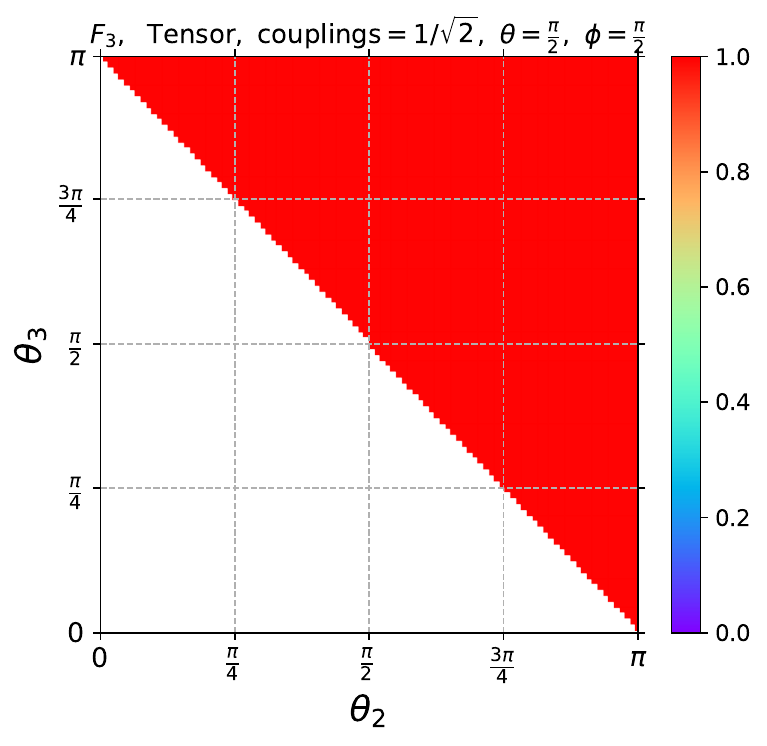}
\includegraphics[scale=.32]
{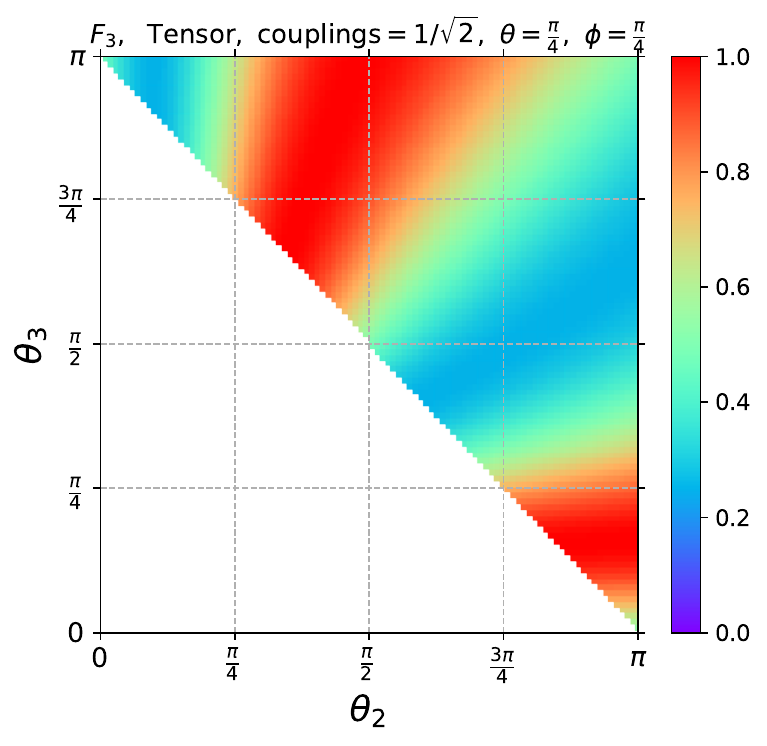}
\caption{\label{fig:2d}
$F_3$ on the ($\theta_2$, $\theta_3$) planes.
}
\end{figure}

In Fig.\ \ref{fig:2d}, we show the GME measure $F_3$ as a function of $\theta_2$ and $\theta_3$ for vector (upper panel) and tensor (lower panel) interactions.
The lower left plane is empty because this region is unphysical, $\theta_2 + \theta_3 < \pi$.
All couplings, $c_X, d_X$ ($X = L,R,M,E$), are fixed to $1/\sqrt{2}$.
In the left panel, the spin direction ${\bf n}$ of the initial particle is set to the $y$ direction (perpendicular to the decay plane), while in the right panel, it is tilted 
with $\theta = \phi = \tfrac{\pi}{4}$ (see Fig.\ \ref{fig:angle}).
% We see that for the vector interaction 
% with $d_L = d_R$,
% $F_3$ is symmetric under $2 \leftrightarrow 3$
% regardless of ${\bf n}$, 
% as can readily be checked from the analytical expressions:
% $M_{LL} \leftrightarrow M_{LR}$
% and 
% $M_{RL} \leftrightarrow M_{RR}$
% under $2 \leftrightarrow 3$.
% We also observe that $F_3$ vanishes at $\theta_2 = \theta_3$, as ${\cal C}_{1(23)}$ vanishes in this configuration.
We see that 
when ${\bf n}$ is perpendicular to the decay plane,
$F_3$ for the vector interaction depends only on the combination $\theta_2 + \theta_3$ and symmetric under $2 \leftrightarrow 3$ exchange.
For the tensor interaction, 
in this case, 
the system is maximally entangled, $F_3 = 1$, regardless of the decay angles.
This results in $|M_L| = |M_R| = 1/\sqrt{2}$, as inferred from Eq.\ \eqref{M_tens}.
When the initial spin is tilted, 
$F_3$ behaves asymmetrically 
under 
$2 \leftrightarrow 3$ 
both for vector and tensor interactions,
as shown in the two right plots of Fig.\ \ref{fig:2d}.

\begin{figure}[t!]
\centering
\includegraphics[scale=.3]{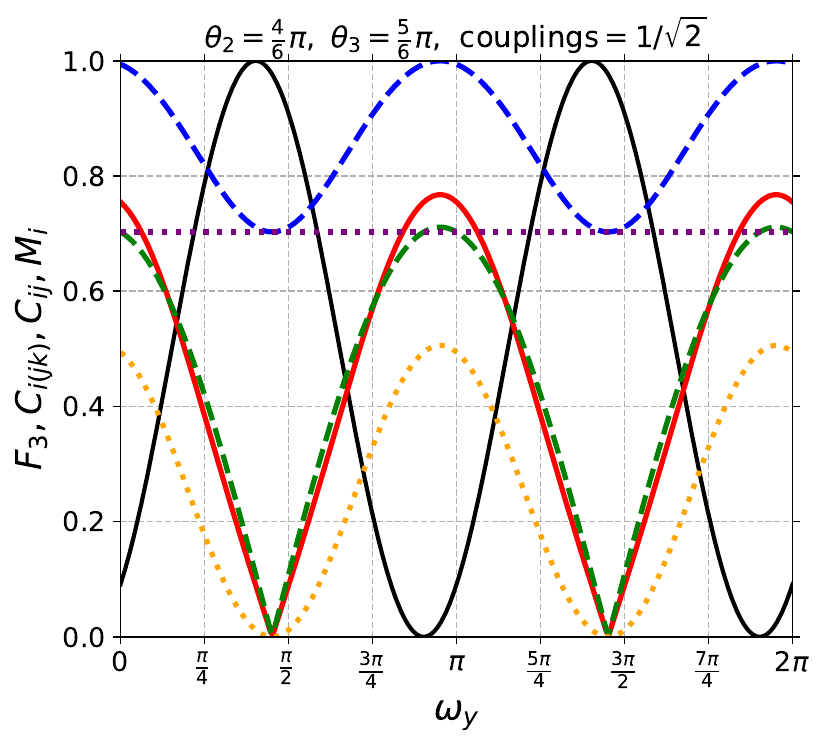}
\includegraphics[scale=.3]{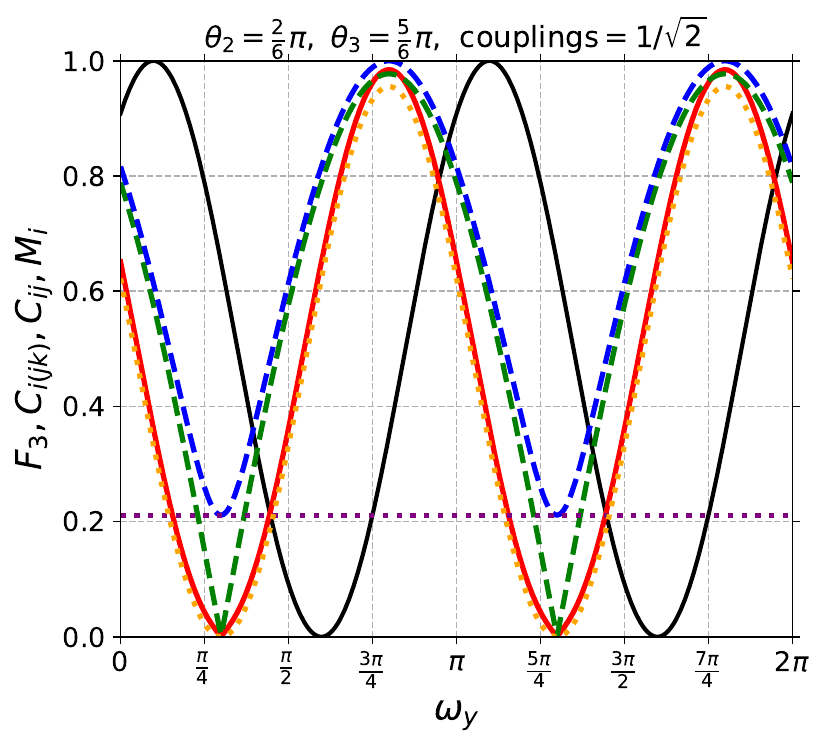}
\includegraphics[scale=.3]{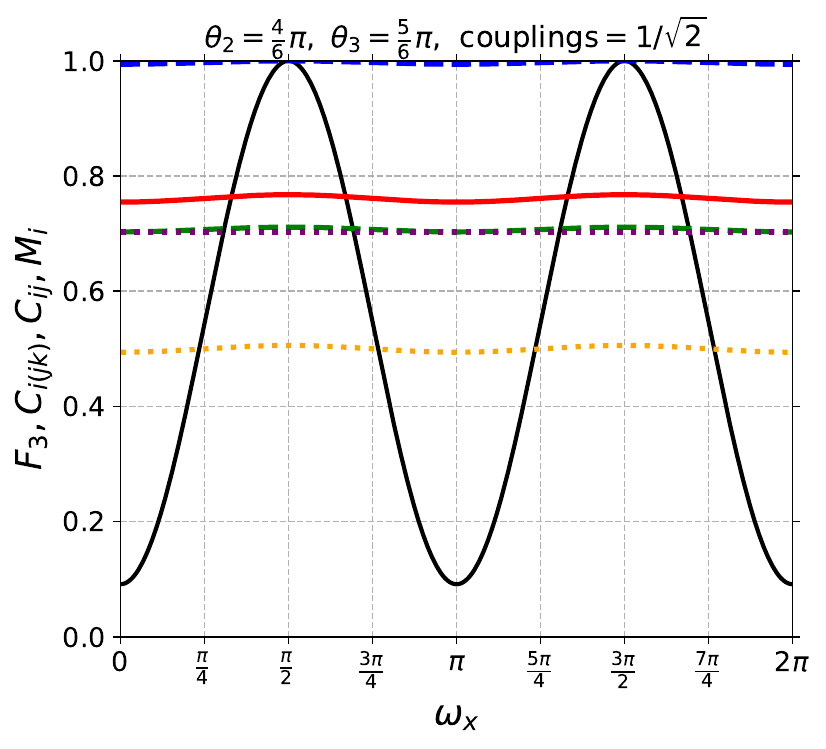}
\includegraphics[scale=.3]{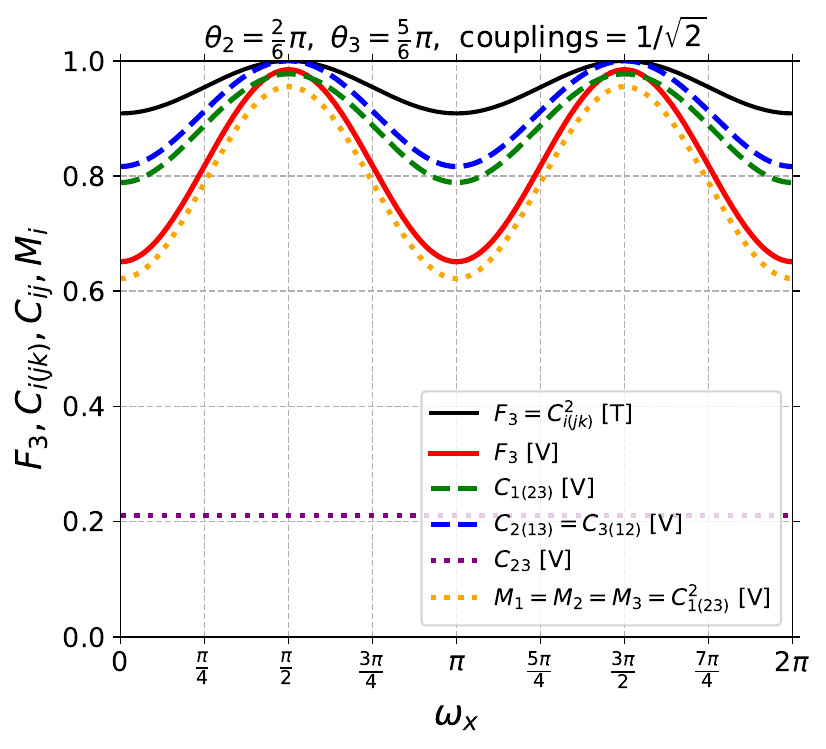}
\caption{\label{fig:1d} 
Various entanglement measures as functions of the initial spin direction ${\bf n}$. }
\end{figure}

Fig.\ \ref{fig:1d} shows various entanglement measures as a function of the initial spin direction {\bf n}.
For vector interaction,
we show 
$F_3$ (red-solid),
${\cal C}_{1(23)}$ (blue-dashed),
${\cal C}_{2(13)}={\cal C}_{3(12)}$ (green-dashed),
${\cal C}_{23}$ (purple-dotted)
and
$M_i = {\cal C}^2_{1(23)}$ (orange-dotted),
while only 
$F_3$ is shown 
for the tensor interaction.
All couplings are set to $1/\sqrt{2}$.
In the right and left panels, the decay angles are fixed to
$(\theta_2, \theta_3) = (\frac{4 \pi}{6}, \frac{5 \pi}{6})$
and
$(\frac{2 \pi}{6}, \frac{5 \pi}{6})$, receptively.
The horizontal axes of the plots represent the angle between
the $z$-axis and ${\bf n}$.
In the upper and lower panels, ${\bf n}$ rotates about the $y$ and $x$ axes 
in the right-handed way, respectively.
We observe that $F_3$ responds differently 
to the rotations of ${\bf n}$
between the vector and tensor cases.
The only non-vanishing 2-particle entanglement ${\cal C}_{23}$ in the vector case is constant with respect to  of ${\bf n}$.

\section{Conclusion}

The exploration of quantum entanglement has been a cornerstone in understanding the non-local correlations inherent in quantum systems. While much of the historical focus has been on two-particle entanglement, we expanded its realm to three-particle systems, revealing a richer tapestry of quantum correlations. This advancement offers profound insights into the fundamental nature of quantum mechanics, extending beyond the traditionally studied bipartite systems.

Building upon the foundational concepts of entanglement monotone concurrence and the monogamy property, we propose a novel approach to understanding entanglement in three-body decays. This exploration paves the way for future studies in multi-particle quantum entanglement and emphasises its significance in particle phenomenology, particularly in the decay dynamics of heavy fermions and hadrons. Having explicitly calculated the expected entanglement for the three-body decay via (pseudo)scalar, (pseudo)vector and (pseudo)tensor mediators, this approach can potentially uncover deviations from the predictions of the Standard Model, shedding light on uncharted territories within the quantum realm.

Thus, we emphasis the pivotal role of three-particle entanglement in particle physics, suggesting new avenues for exploring new observables and novel search strategies in high-energy physics.

\section{Acknowledgements}

The work of K.S.\ is partially supported by the National Science Centre, Poland, under research grant 2017/26/E/ST2/00135
and
has received 
funding from the Norwegian Financial Mechanism for years 2014-2021, 
grant nr DEC-2019/34/H/ST2/00707. 
M.S.\ is supported by the STFC under grant ST/P001246/1

\bibliographystyle{apsrev4-1}

%\bibliography{bibLibrary}  % bibtex entries in separate file for convenience
\bibliography{refs}

\end{document}